\documentclass[aps,prl,twocolumn,groupedaddress,showpacs,showkeys,superscriptaddress,nofootinbib]{revtex4}%
\usepackage{mathrsfs} 
\begin{document}%
\title{Phenomenologically viable Lorentz-violating quantum gravity}
\author{Thomas P. Sotiriou}
\affiliation{Center for Fundamental Physics, University of Maryland, College Park, MD20742-4111, USA }
\author{Matt Visser}
\affiliation{School of Mathematics, Statistics, and Operations Research,
Victoria University of Wellington, New Zealand}
\author{Silke Weinfurtner}
\affiliation{Physics and Astronomy Department, University of British Columbia, Vancouver, Canada}
\date{28 April 2009; 28 May 2009; 
\LaTeX-ed \today}
\bigskip
\begin{abstract}
Ho\v{r}ava's ``Lifschitz point gravity'' has many desirable features, but in its original incarnation one is forced to accept a non-zero cosmological constant of the wrong sign to be compatible with observation. We develop an extension of Ho\v{r}ava's model that abandons ``detailed balance'' and regains parity invariance, and in 3+1 dimensions exhibit all five marginal (renormalizable) and four relevant (super-renormalizable) operators, as determined by power counting. We also consider the classical limit of this theory, evaluate the Hamiltonian and super-momentum constraints, and extract the classical equations of motion in a form similar to the ADM formulation of general relativity. This puts the model in a framework amenable to developing detailed precision tests.

\end{abstract}
\keywords{quantum gravity; anisotropic scaling; Lifschitz point\\
arXiv:0904.4464 [hep-th]. Published as Physical Review Letters {\bf 102} (2009) 251601}
\maketitle

\newcommand{\scri}{\mathscr{I}}
\newcommand{\sun}{\ensuremath{\odot}}%
\def\e{{\mathrm e}}%
\def\g{{\mbox{\sl g}}}%
\def\Box{\nabla^2}%
\def\d{{\mathrm d}}%
\def\R{{\rm I\!R}}%
\def\ie{{\em i.e.\/}}%
\def\eg{{\em e.g.\/}}%
\def\etc{{\em etc.\/}}%
\def\etal{{\em et al.\/}}%
\def\HRULE{{\bigskip\hrule\bigskip}}

\def\d{{\mathrm{d}}}
\def\J{{\mathscr{J}}}
\def\L{{\mathscr{L}}}
\def\H{{\mathscr{H}}}
\def\T{{\mathscr{T}}}
\def\V{{\mathscr{V}}}

\def\sech{{\mathrm{sech}}}

Quantum gravity models based on the ``anisotropic scaling'' of space and time have recently attracted significant attention~\cite{Horava, Pal, LSB, cosmology, Natase}. In particular, Ho\v{r}ava's  ``Lifschitz point gravity''~\cite{Horava} has very many desirable features, but in its original incarnation one is forced to accept a non-zero cosmological constant of the wrong sign to be compatible with observation (see also~\cite{Natase}). We outline a variant of Ho\v{r}ava's model that appears to be phenomenologically viable. 

The basic idea~\cite{Horava} is to write the spacetime metric in ADM form
\begin{equation}
\d s^2 = - N^2 c^2 \d t^2 + g_{ij}(\d x^i - N^i \d t) (\d x^j - N^j \d t),
\end{equation}
and then, (adopting $\kappa$ as a placeholder symbol for some object with the dimensions of momentum), postulate that the engineering dimensions of  space and time are
\begin{equation}
[\d x] = [\kappa]^{-1}; \qquad [\d t] = [\kappa]^{-z}. 
\end{equation}
Effectively, one is implicitly introducing a quantity $Z$, with the physical dimensions $[Z]=[\d x]^z/[\d t]$, and using the theorists' prerogative to adopt units such that $Z\to1$.  (Ultimately we shall interpret this quantity $Z$ in terms of the Planck scale, and closely related Lorentz-symmetry breaking scales.) If one prefers to characterize this quantity $Z$  in terms of a momentum $\zeta$, then $Z=\zeta^{-z+1} c$ and we see that in order for dimensional analysis to be useful one cannot simultaneously set both $Z\to1$ and $c\to1$.  (Attempting to do so forces both $\d x$ and $\d t$ to be dimensionless, which then renders dimensional analysis impotent.) 
Consequently, in theoretician's ($Z\to1$) units  one \emph{must} have
\begin{equation}
[N^i] = [c]  = {[\d x]\over[\d t]} = [\kappa]^{z-1},
\end{equation}
and one is free to additionally choose
\begin{equation}
[g_{ij}]  = [N] =  [1];\qquad  [\d s] = [\kappa]^{-1}.
\end{equation}
To minimize the algebraic manipulations, it is further convenient to take the volume element to be
\begin{equation}
\d V_{d+1} =  \sqrt{g}\,  N \; \d^d x \; \d t; \qquad [\d V_{d+1}] = [\kappa]^{-d-z}.
\end{equation}
Note the \emph{absence} of any factor of $c$ in this definition. The resulting model will, by its very construction, violate Lorentz invariance;  the payoff however is greatly improved ultraviolet (UV) behaviour for the Feynman diagrams~\cite{Horava, Pal, LSB}, coupled with a well-behaved low-energy infrared (IR) limit.

We shall argue that an extension of the specific model presented in~\cite{Horava} is (at least superficially) phenomenologically viable, and has a classical limit that is amenable to analysis in an ADM-like manner. This is one of very few quantum gravity models that has any realistic hope of direct confrontation with experiment and observation.

The extrinsic curvature is most conveniently defined to not include any explicit factor of $c$:
\begin{equation}
K_{ij} = {1\over2N} \left\{ - \dot g_{ij} + \nabla_i N_j + \nabla_j N_i \right\}.
\end{equation}
Then $[N^i] =  {[\d x] / [\d t]} = [\kappa]^{z-1}$,
in agreement with the above.  Furthermore
\begin{equation}
[K_{ij}] = {[g_{ij}]\over[N]  [\d t]} = [\kappa]^{z}.
\end{equation}
For the spatial slices we have
\begin{equation}
[g_{ij}] = [1]; \qquad [\Gamma^i{}_{jk}] = [\kappa];  \qquad [R^i{}_{jkl}] = [\kappa]^2,
\end{equation}
the key point being
\begin{equation}
[R^{ijkl}] = [\kappa]^{2};\quad [\nabla R^{ijkl}] = [\kappa]^{3};\quad [\nabla^2 R^{ijkl}] = [\kappa]^{4}.
\end{equation}
For the kinetic term we consider
\begin{equation}
\T(K) = {g_K}     \left\{ (K^{ij} K_{ij} - K^2)+ \xi K^2 \right\}.
\end{equation}
(Standard general relativity would enforce $\xi\to0$.)
Take the kinetic action to be
\begin{equation}
S_K =    \int \T(K) \; \d V_{d+1} =    \int \T(K) \sqrt{g} \; N\; \d^d x \; \d t.
\end{equation}
Again, note that for convenience it is most useful to arrange the \emph{absence} of any explicit factors of $c$.
Then 
\begin{equation}
[S_K]  =   [g_K]  [\kappa]^{z-d}.
\end{equation}
But since the kinetic action is (by definition) chosen to be dimensionless, we have
\begin{equation}
[g_K] =  [\kappa]^{(d-z)}.
\end{equation} 
Note that the kinetic coupling constant $g_K$ is dimensionless exactly for $d=z$, 
which is exactly the condition that was aimed for in~\cite{Horava}. In a simplified model based on scalar field self interactions, this is exactly the condition for well-behaved UV behaviour derived in~\cite{LSB}. Once we have set $d=z$ to make $g_K$ dimensionless, then provided $g_K$ is positive one can without loss of generality re-scale the time and/or space coordinates to set $g_K \to 1$.

Now consider the potential term
\begin{equation}
S_\V =  -\int   \V(g) \; \d V_{d+1} = - \int   \V(g) \; \sqrt{g} \; N \; \d^d x \; \d t,
\end{equation}
where $\V(g)$ is some scalar built out of the spatial metric and its derivatives.
Then 
\begin{equation}
[S_\V] =  [\V(g)]  [\kappa]^{-d-z},
\end{equation}
whence
\begin{equation}
 [\V(g)] \to [\kappa]^{d+z}.
 \end{equation}
But to keep the kinetic coupling $g_K$ dimensionless we needed
$z\to d$. Therefore 
\begin{equation}
 [\V(g)] \to [\kappa]^{2d}.
 \end{equation}
But $\V(g)$ must be built out of scalar invariants calculable in terms of the Riemann tensor (Rm) and its derivatives, which tells us it must be constructible from objects of the form
\begin{equation}
\left\{  (\hbox{Rm})^d,  [(\nabla\hbox{Rm})]^2 \hbox{(Rm})^{d-3} , \hbox{etc...}   \right\}.
\end{equation}
In general, in $d+1$ dimensions this is a long but finite list. All of these theories should be well-behaved as quantum field theories~\cite{Horava, LSB}.
(In particular,  certain key aspects of~\cite{Horava}  generalize nicely to $d+1$ dimensions.)
In the specific case $d=3$ we have
\begin{equation}
 [\V(g)] \to [\kappa]^{6},
 \end{equation}
 and so obtain the short and rather specific list:
\begin{equation}
\Big\{  (\hbox{Rm})^3,  [\nabla(\hbox{Rm})]^2,    (\hbox{Rm})  \nabla^2(\hbox{Rm}),   \nabla^4(\hbox{Rm}) \Big\}.
\end{equation}
But in 3 dimensions the Weyl tensor automatically vanishes, so we can always decompose the Riemann tensor into the Ricci tensor (Rc), Ricci scalar, plus the metric. Thus we need only consider the much simplified list:
\begin{equation}
\Big\{  (\hbox{Rc})^3,  [\nabla(\hbox{Rc})]^2,  (\hbox{Rc})  \nabla^2(\hbox{Rc}),   \nabla^4(\hbox{Rc}) \Big\}. \quad
\end{equation}
We now consider a model that generalizes that of Ho\v{r}ava~\cite{Horava} by containing all possible terms of this type, eliminating redundant terms using:
(i) Integration by parts and discarding surface terms;
(ii) Commutator identities;
(iii) Bianchi identities;
(iv) Special relations appropriate to 3 dimensions. (Weyl vanishes; properties of  Cotton tensor.)
We do not need explicit parity violation, and for simplicity we choose to exclude it.

Ho\v{r}ava's  prescription for keeping the calculation tractable was to impose two simplifications: (i) a ``projectability condition'' on the lapse function~\cite{Horava}, (this effectively is the demand that the lapse $N(t)$ is a function of $t$ only, not a function of position), and (ii) a condition Ho\v{r}ava referred to as ``detailed balance''~\cite{Horava}. We shall retain the ``projectability condition'' but abandon  ``detailed balance''. We consider ``detailed balance'' to be too restrictive and physically unnecessary.

It should be remarked that in standard general relativity the ``projectability condition''  can always be enforced locally as a gauge choice; furthermore for physically interesting solutions of general relativity (though not necessarily for perturbations around those solutions) it seems that this can always be done globally. For instance, for the Schwarzschild spacetime this ``projectability condition'' holds globally in Painlev\'e--Gullstrand coordinates~\cite{Living}, while in the Kerr spacetime this condition holds globally (for the physically interesting  $r>0$ region) in Doran coordinates~\cite{Visser-on-Kerr}. Furthermore FLRW cosmologies also automatically satisfy this ``projectability condition''. 

However, there is a  price to pay for enforcing it at the level of the action (and before any functional variation): the theory we are considering is not necessarily the most general theory with all possible terms of dimension six. (But it is still general enough to be a significant generalization with respect to Ho\v{r}ava's model~\cite{Horava}).

After a brief calculation, we find that there are only five independent terms of dimension $[\kappa]^6$:
\begin{equation}
 R^3, \quad R R^{i}{}_{j} R^{j}{}_i, \quad  R^i{}_j R^j{}_k R^k{}_i; \quad 
R \; \nabla^2 R, \quad  \nabla_i R_{jk} \, \nabla^i R^{jk}.
\end{equation}
These terms are all marginal (renormalizable) by power counting~\cite{Horava, LSB}. In Ho\v{r}ava's article~\cite{Horava} only a particular linear combination of these five terms is considered: Phrased in terms of the Cotton tensor, Ho\v{r}ava considers the single $[\kappa]^6$ term $C^{i}{}_{j}  C^{j}{}_{i}$. 

If we now additionally add all possible lower-dimension terms (relevant operators, super-renormalizable by power-counting) we obtain four additional operators:
\begin{equation}
[\kappa]^0: \; \;  1;
\qquad
[\kappa]^2: \; \;   R;
\qquad
[\kappa]^4: \; \;   R^2; \; \; R^{ij} R_{ij}.
\end{equation}
This now results in a potential $\V(g)$ with nine terms and nine independent coupling constants. 
In contrast, Ho\v{r}ava~\cite{Horava} chooses a potential $\V(g)$ containing six terms with only three independent coupling constants, of the form $(\tilde g_2\,\hbox{Cotton}+\tilde g_1\,\hbox{Einstein}+\tilde g_0\,\hbox{metric})^2$.

Assembling all the pieces we now have
\begin{equation}
S =  \int  \left[  \T(K) -  \V(g)  \right]  \sqrt{g}  \; N\; \d^3 x \; \d t,
\end{equation}
with
\begin{eqnarray}
\V(g)  &=&   g_0\, \zeta^6 +  g_1\, \zeta^4 \, R + g_2 \,\zeta^2\, R^2 + g_3 \, \zeta^2\,R_{ij} R^{ij} 
\nonumber\\
&&
+ g_4 \,R^3 + g_5 \,R (R_{ij} R^{ij}) + g_6\, R^i{}_j R^j{}_k R^k{}_i 
\nonumber\\
&&+ g_7 \,R \nabla^2 R + g_8\,  \nabla_i R_{jk} \, \nabla^i R^{jk},
\end{eqnarray}
where we have introduced suitable factors of $\zeta$ to ensure the couplings $g_a$ are all dimensionless.
Now assuming $g_1<0$, we can without loss of generality re-scale the time and space coordinates to set \emph{both} $g_K \to 1$ and $g_1\to -1$.
The Einstein--Hilbert piece of the action is now
\begin{equation}
S_\mathrm{EH} =   \int   \left\{  (K^{ij} K_{ij} - K^2) +  \zeta^4 R  - g_0\, \zeta^6 \right\} \sqrt{g} \; N\; \d^3 x \; \d t,
\end{equation}
and the ``extra'' Lorentz-violating terms are
\begin{eqnarray}
S_\mathrm{LV} &=&  \int   \Big\{ \xi\, K^2 -     g_2 \,\zeta^2\,R^2 -  g_3 \, \zeta^2\, R_{ij} R^{ij} 
\nonumber\\
&&
- g_4 \, R^3 - g_5 \, R (R_{ij} R^{ij}) - g_6 \, R^i{}_j R^j{}_k R^k{}_i 
\nonumber\\
&&
- g_7 \, R \nabla^2 R - g_8 \, \nabla_i R_{jk} \, \nabla^i R^{jk}
\Big\} \sqrt{g} \; N\; \d^d x \; \d t. \qquad
\end{eqnarray}
This perfectly reasonable classical Lorentz-violating theory of gravity certainly deserves study in its own right. 

While these $Z\to1$ units have been most useful for power counting purposes, when it comes to phenomenological confrontation with observation it is much more useful to adapt more standard ``physical'' ($c\to1$) units, in which $[\d x]=[\d t]$. The transformation to physical units is most easily accomplished by setting $(\d t)_{Z=1} \to \zeta^{-2} (\d t)_{c=1}$.  In these physical units the Einstein--Hilbert piece of the action becomes
\begin{equation}
S_\mathrm{EH} =   \zeta^2 \int   \left\{  (K^{ij} K_{ij} - K^2) +  R  -  g_0\, \zeta^2 \right\} \sqrt{g} \; N\; \d^3 x \; \d t,
\end{equation}
and the ``extra'' Lorentz-violating terms become
\begin{eqnarray}
S_\mathrm{LV} &=&  \zeta^2 \int   \Big\{ \xi\, K^2 -     g_2 \,\zeta^{-2}\,R^2 -  g_3 \, \zeta^{-2}\, R_{ij} R^{ij} 
\nonumber\\
&&
- g_4 \,  \zeta^{-4}\,R^3 - g_5 \,\zeta^{-4}\, R (R_{ij} R^{ij})
\nonumber\\
&&
- g_6 \,\zeta^{-4}\, R^i{}_j R^j{}_k R^k{}_i 
- g_7\,\zeta^{-4}\, R \nabla^2 R
\nonumber\\
&&
 - g_8 \,\zeta^{-4}\, \nabla_i R_{jk} \, \nabla^i R^{jk}
\Big\} \sqrt{g} \; N\; \d^d x \; \d t. \qquad
\end{eqnarray}
From this normalization of the Einstein--Hilbert term, we see that in physical ($c\to1$) units
\begin{equation}
(16 \pi G_\mathrm{Newton})^{-1} = \zeta^2;  \quad \Lambda  =  {g_0 \, \zeta^2\over2};
\end{equation}
so that $\zeta$ is identified as the Planck scale.  The cosmological constant is determined by the free parameter $g_0$, and observationally $g_0 \sim 10^{-123}$ (renormalized after including any vacuum energy contributions). In particular, the way we have set this up we are free to choose the Newton constant and cosmological constant independently (and so to be compatible with observation). In contrast, in the original model presented in~\cite{Horava}, a non-zero Newton constant \emph{requires} a non-zero cosmological constant, and as long as Ho\v{r}ava's ``detailed balance'' symmetry is preserved, this will be of the \emph{wrong sign} to be compatible with cosmological observations.  

The extra Lorentz violating terms consist of one kinetic and seven higher-curvature terms.
The Lorentz violating term in the kinetic energy leads to an extra scalar mode for the graviton~\cite{Horava}, with fractional $O(\xi)$ effects at all momenta.  Phenomenologically, this behaviour is potentially dangerous and should be carefully investigated. 
In contrast the  various Lorentz-violating terms in the potential become comparable to the spatial curvature term in the Einstein--Hilbert action for physical momenta of order
\begin{equation}
\zeta_{\{2,3\}} = {\zeta\over\sqrt{ |g_{\{2,3\}}| }};  \quad \zeta_{\{4,5,6,7,8\}} = {\zeta\over\sqrt[4]{ |g_{\{4,5,6,7,8\}}| }}.
\end{equation}
Thus the higher-curvature terms are automatically suppressed as we go to low curvature (low momentum).  Note that we have also divorced the Planck scale $\zeta$ from the various Lorentz-breaking scales $\zeta_{\{2,3,4,5,6,7,8\}}$, and that we can drive the Lorentz breaking scale arbitrarily high by suitable adjustment of the dimensionless couplings $ g_{\{2,3\}}$ and  $g_{\{4,5,6,7,8\}}$. It is these pleasant properties that make the model phenomenologically viable --- at least at a superficial level --- and that encourage us to consider more detailed confrontation with experiment and observation.  Since the UV dominant part of the Lorentz breaking is sixth order in momenta, in the absence of significant UV--IR mixing it neatly evades all current bounds  on Lorentz symmetry breaking~\cite{Mattingly, Liberati, Planck}. The potentially risky issue of UV--IR mixing should also be carefully investigated in this model~\cite{UV-IR}.  That UV--IR mixing is not invariably fatal can be inferred from the fact that \emph{observationally} many condensed-matter analogue systems exhibit emergent Lorentz symmetry in the IR~\cite{Living}, and that certain specific systems exhibit a ``natural'' suppression of Lorentz violating effects~\cite{naturalness}.

Varying with respect to the lapse $N(t)$ one obtains the Hamiltionian constraint
\begin{equation}
H = \int \sqrt{g} \,\H \,\d^3 x =\int \sqrt{g} \,\{ \T(K) + \V(g) \} \, \d^3 x = 0.
\end{equation}
The difference compared to standard general relativity lies in (i) the $\xi$ term in the kinetic energy, (ii) the more complicated form of the potential $\V(g)$, and (iii) because of the assumed ``projectability condition'' on the lapse $N(t)$ one cannot derive a super-Hamiltonian constraint, and must remain satisfied with this \emph{spatially integrated} Hamiltonian constraint.

Varying with respect to the shift $N^i$ one obtains the super-momentum constraint 
\begin{equation}
\nabla_i \pi^{ij}= 0,
\end{equation}
where the super-momentum is
\begin{equation}
\pi^{ij} = {\partial [N \,\T(K)] \over\partial \dot g_{ij} } = - \left\{ K^{ij} - K g^{ij}  + \xi K g^{ij} \right\}.
\end{equation}
The difference compared to standard general relativity is utterly minimal and lies solely in the $\xi$ term.

By varying with respect to $g_{ij}$ one now obtains the dynamical equation
\begin{eqnarray}
{1 \over \sqrt{g}} \; \partial_t ( \sqrt{g} \; \pi^{ij} ) &=&  -2N \left\{ (K^2)^{ij} - K K^{ij} + \xi K K^{ij} \right\} 
\nonumber\\
&& + {N\over2} \T(K) \; g^{ij}  + (\nabla_m N^m) \; \pi^{ij} 
\nonumber\\
&& + [\L_{\vec N} \pi]^{ij}
+{N\over \sqrt{g}}   {\delta S_\V \over\delta g_{ij} }.
\end{eqnarray}
This is very similar to standard general relativity: There is a straightforward extra contribution coming from the $\xi$ term in the kinetic energy, but the only real subtlety lies in evaluating the ${\delta S_\V /\delta g_{ij} }$ terms. This is somewhat tedious, but since we know that $S_\V$ is the most general action one can build out of the metric using 0, 2, 4, or 6 derivatives we can 
deduce that the ``forcing term''
\begin{equation}
F^{ij} = {1\over \sqrt{g}}   {\delta S_\V \over\delta g_{ij} }
\end{equation}
is the most general symmetric conserved tensor one can build out of the metric and 0, 2, 4, or 6 derivatives. (An explicit evaluation of these terms has been performed, but the result is too long to write down here, details are provided elsewhere~\cite{elsewhere}.) 
The relevance of these observations is that the classical limit has now been cast into an ADM-like form, suitable, for instance, for detailed numerical investigations.

The model so far only considers pure gravity, and seems to be very well-behaved. It is a very definite proposal with a small number of adjustable parameters, (many fewer adjustable parameters than the standard model of particle physics), making it worthwhile to put in the additional effort to develop precision tests that would confront this model with experimental and observational bounds.  The most obvious tests would come from the observational limits on Lorentz violations~\cite{Mattingly, Liberati, Planck, UV-IR}.  By inspection the model should also fall into the PPN framework, and specifically be subject to ``preferred frame'' effects~\cite{PPN} --- this should lead to stringent limits on the size of the Lorentz breaking parameters $\zeta_a$ arising from solar system physics.  Up to this stage we have not had to make any specific commitment as to how matter couples to the gravitational field: this is a key open problem for future investigations.

In conclusion, while there is certainly a tremendous amount of work still to be done, we would argue that this model could very well in its own right be a promising candidate for quantum gravity, or alternatively for the effective field theory resulting from some more fundamental theory. Last but certainly not least, the model discussed above is one of very few quantum gravity models that has any realistic hope  of direct confrontation with experiment  and observation, and so is well worth a very careful look.

TPS was supported by NSF grant PHYS-0601800. MV was supported by the NZ Marsden Fund. SW was supported by a Marie Curie Fellowship.

\vskip - 15pt


\begin{thebibliography}{99}
\bibitem{Horava}
\vskip-10pt
P.~Ho\v{r}ava,
  Phys.\ Rev.\  D {\bf 79}, 084008 (2009) 
  [arXiv:0901.3775 [hep-th]],
  JHEP 0903:020,2009
  [arXiv:0812.4287 [hep-th]].
  Phys. Rev. Lett. {\bf 102} (2009) 161301
  [arXiv:0902.3657 [hep-th]].

\bibitem{Pal}
S.~Pal,
  arXiv:0901.0599 [hep-th].
  
 \bibitem{LSB}
M.~Visser,
  arXiv:0902.0590 [hep-th].
  
 \bibitem{cosmology}
  T.~Takahashi and J.~Soda,
  arXiv:0904.0554 [hep-th] \,\,[Phys. Rev. Lett. (to be published)].
    G.~Calcagni,
  arXiv:0904.0829 [hep-th].
  E.~Kiritsis and G.~Kofinas,
  arXiv:0904.1334 [hep-th].
  H.~Lu, J.~Mei and C.~N.~Pope,
  arXiv:0904.1595 [hep-th].
  S.~Mukohyama,
  arXiv:0904.2190 [hep-th].
 R.~G.~Cai, L.~M.~Cao and N.~Ohta,
arXiv:0904.3670 [hep-th].
  
  \bibitem{Natase}
  H.~Nastase,
  arXiv:0904.3604 [hep-th].
  
  \bibitem{Living}
  C.~Barcel\'o, S.~Liberati and M.~Visser,
  Living Rev.\ Rel.\  {\bf 8}, 12 (2005)
  [arXiv:gr-qc/0505065].
   
  \bibitem{Visser-on-Kerr}
  M.~Visser,
  ``The Kerr spacetime: A brief introduction,''
  arXiv:0706.0622 [gr-qc]. 
  Published in \emph{The Kerr spacetime: Rotating black holes in general relativity}, edited by D.~Wiltshire, M.~Visser, and S.~Scott, (Cambridge University Press, 2009).
  
   \bibitem{Mattingly}
 D.~Mattingly,
  Living Rev.\ Rel.\  {\bf 8} (2005) 5
  [arXiv:gr-qc/0502097].
  
 
  \bibitem{Liberati}
  T.~Jacobson, S.~Liberati and D.~Mattingly,
  Annals Phys.\  {\bf 321} (2006) 150
  [arXiv:astro-ph/0505267].
 T.~A.~Jacobson, S.~Liberati, D.~Mattingly and F.~W.~Stecker,
  Phys.\ Rev.\ Lett.\  {\bf 93} (2004) 021101
  [arXiv:astro-ph/0309681].
  T.~Jacobson, S.~Liberati and D.~Mattingly,
  Nature {\bf 424} (2003) 1019
  [arXiv:astro-ph/0212190].
  

  \bibitem{Planck}
L.~Maccione, A.~M.~Taylor, D.~M.~Mattingly and S.~Liberati,
  arXiv:0902.1756 [astro-ph.HE].
  
 
  
   \bibitem{UV-IR}
R.~C.~Myers and M.~Pospelov,
  Phys.\ Rev.\ Lett.\  {\bf 90} (2003) 211601
  [arXiv:hep-ph/0301124].
J.~Collins, A.~Perez, D.~Sudarsky, L.~Urrutia and H.~Vucetich,
  Phys.\ Rev.\ Lett.\  {\bf 93} (2004) 191301
  [arXiv:gr-qc/0403053].
  O.~Gagnon and G.~D.~Moore,
  Phys.\ Rev.\  D {\bf 70} (2004) 065002
  [arXiv:hep-ph/0404196].

\bibitem{naturalness}
S.~Liberati, M.~Visser and S.~Weinfurtner,
  Phys.\ Rev.\ Lett.\  {\bf 96}, 151301 (2006)
  [arXiv:\-gr-qc/0512139].
  
\bibitem{elsewhere}
T.~P.~Sotiriou, M.~Visser and S.~Weinfurtner,
  arXiv:0905.2798 [hep-th].
  
  \enlargethispage{45pt}

 \bibitem{PPN}
 C.~M.~Will and K.~J.~Nordtvedt,
  Astrophys.\ J.\  {\bf 177} (1972) 757.
 



\end{thebibliography}
\end{document}